\documentclass[12pt]{article}
\usepackage{amsmath,amssymb,amsthm}
\usepackage[dvips]{graphicx}
\theoremstyle{definition}

\theoremstyle{remark}

\newcommand{\beq}{\begin{eqnarray}}
\newcommand{\eeq}{\end{eqnarray}}
\newcommand{\beqnn}{\begin{eqnarray*}}
\newcommand{\eeqnn}{\end{eqnarray*}}
\newcommand{\rd}{\partial}

\newcommand{\tp}[1]{\:{}^{\mathrm{t}}#1}
\newcommand{\CC}{\mathbf{C}}
\newcommand{\PP}{\mathbf{P}}

\newcommand{\bst}{\boldsymbol{t}}
\newcommand{\bsx}{\boldsymbol{x}}
\newcommand{\bsy}{\boldsymbol{y}}
\newcommand{\bsz}{\boldsymbol{z}}
\newcommand{\bszero}{\boldsymbol{0}}
\newcommand{\calO}{\mathcal{O}}
\newcommand{\calP}{\mathcal{P}}
\newcommand{\SSTab}{\mathrm{SSTab}}

\begin{document}

\title{Remarks on partition functions of \\
topological string theory on \\
generalized conifolds
\footnote{Contribution to the proceedings of the RIMS camp-style seminar 
``Algebraic combinatorics related to Young diagrams and statistical physics'',  
August, 2012, International Institute for Advanced Studies, Kyoto, 
organized by M. Ishikawa, S. Okada and H. Tagawa. } 
}
\author{Kanehisa Takasaki\\
{\normalsize Graduate School of Human and Environmental Studies, 
Kyoto University}\\
{\normalsize Yoshida, Sakyo, Kyoto 606-8501, Japan}\\
{\normalsize takasaki@math.h.kyoto-u.ac.jp} }
\date{}
\maketitle

\begin{abstract}
The method of topological vertex for topological string theory 
on toric Calabi-Yau 3-folds is reviewed.  Implications of 
an explicit formula of partition functions in the ``on-strip'' case, 
typically the generalized conifolds, are considered.  
Generating functions of part of the partition functions are 
shown to be tau functions of the KP hierarchy.  The associated 
Baker-Akhiezer functions play the role of wave functions, and 
satisfy $q$-difference equations.  These $q$-difference equations 
represent the quantum mirror curves conjectured by Gukov and 
Su{\l}kowski.  
\end{abstract}

\begin{flushleft}
\end{flushleft}

\section{Introduction}

Topological string theory is a simplified version 
of string theory in which topological properties 
of the target space are captured by dynamics of strings 
\cite{Marino-book}.  Some ten years ago, the method 
of ``topological vertex'' was introduced as a technique 
for calculating the ``amplitudes'' of topological string theory 
on toric (or, more precisely, local toric) 
Calabi-Yau 3-folds \cite{AKMV03}.  
This method enables one to describe the amplitudes 
in the language of purely combinatorial notions 
such as partitions and (skew) Schur functions.  

Toric Calabi-Yau 3-folds are characterized 
by graphical data called ``web diagrams'' 
(which are dual to two-dimensional ``toric diagrams'').  
The vertices of a web diagram represent copies 
of the simplest Calabi-Yau 3-fold $\CC^3$.  
These $\CC^3$'s are glued together to form 
the 3-fold $X$ in question.  The web (or toric) diagram 
encodes the gluing data.  

The topological vertex is the amplitude $C_{\alpha\beta\gamma}$ 
of topological string theory on $\CC^3$ 
with boundary conditions imposed on string world sheets.  
The indices $\alpha,\beta,\gamma$ are integer partitions 
that specify the boundary conditions.  
These partitions are placed on the three edges 
emanating from the vertex.  When the copies of $\CC^3$ 
are glued together, their vertex weights 
are multiplied along with edges weights, 
and summed over all possible configurations 
of the partitions on the internal edges.  
The vertex weight $C_{\alpha\beta\gamma}$ itself 
is a somewhat complicated combination of special values 
of skew Schur functions.  Thus the amplitude 
of topological string theory on $X$, also called 
the ``partition function'' from the point of view of 
statistical mechanics, is a sum of combinatorial quantities 
with respect to the partitions on the inner edges.  

When the toric diagram is a triangulated strip, 
e.g., in the case of the resolved conifold 
and its generalizations, one can calculate 
the sum over partitions explicitly with the aid 
of the Cauchy identities for skew Schur functions.  
In this paper, we review this result and consider its implications 
in the context of ``integrable hierarchies'' \cite{ADKMV03,Zhou0310b,DZ12}
and ``quantum mirror curves'' \cite{GS11,Zhou12}.  In particular, 
we present an explicit form of $q$-difference equations 
for ``wave functions'' \cite{KP06} of the generalized conifold. 
This is a generalization of the known result on the resolved conifold.  
Although free fermions and vertex operators are very convenient tools 
\cite{ORV03,BY08}, we dare not use them and resort to the Cauchy identities.

\section{Partitions and Schur functions}

In this section, we recall some relevant notions from combinatorics 
of integer partitions and Schur functions \cite{Macdonald-book}.

\subsection{Partitions}

In this paper, partitions are understood to be 
decreasing sequence of non-negative integers $\lambda_i$, 
$k = 1,2,\ldots$, in which only a finite number are non-zero: 
\beqnn
  \lambda = (\lambda_1,\lambda_2,\ldots), \quad 
  \lambda_1 \ge \lambda_2 \ge \cdots \ge 0,\quad 
  \exists N\; \forall i > N\; \lambda_i = 0. 
\eeqnn
The standard notations 
\beqnn
%\begin{gathered}
  l(\lambda) = \max\{i \mid \lambda_i \not= 0\}, \quad 
  |\lambda| = \sum_{i\ge 1}\lambda_i, \quad %\\
  \kappa(\lambda) = \sum_{i\ge 1}\lambda_i(\lambda_i-2i+1) 
%\end{gathered}
\eeqnn
for the length, the weight and the second Casimir value 
are used throughout this paper. 
Partitions are in one-to-one correspondence with Young diagrams 
by identifying the parts $\lambda_i$, $i = 1,2,\ldots$ 
with the lengths of rows.  Let $\tp{\lambda}$ 
denote the conjugate partition of $\lambda$. 
The parts $\tp{\lambda}_j$, $j = 1,2,\ldots$, of $\tp{\lambda}$ 
are the lengths of columns of the same Young diagram.  
The length $h(i,j)$ of the hook cornered at $(i,j) \in \lambda$ 
can be thereby expressed as 
\beqnn
  h(i,j) = \lambda_i -i + \tp{\lambda}_j-j + 1. 
\eeqnn

Two partitions $\lambda,\mu$ are said to satisfy 
the inclusion relation $\lambda \supset \mu$ 
if the corresponding Young diagrams satisfy 
the inclusion relation (equivalently, 
if their parts $\lambda_i,\mu_i$ satisfy 
the inequalities $\lambda_i \ge \mu_i$ for $i \ge 1$).  
The difference of those Young diagram is denoted 
by $\lambda/\mu$ and called a skew Young diagram 
of shape $\lambda/\mu$.

\subsection{Schur and skew Schur functions}

These partitions label the Schur functions $s_\lambda(\bsx)$
and the skew Schur functions $s_{\lambda/\mu}(\bsx)$  
of a finite or infinite number of variables $\bsx = (x_1,x_2,\cdots)$.  
In a combinatorial definition, $s_\lambda(\bsx)$ is a sum 
of the form 
\beq
  s_\lambda(\bsx) = \sum_{T\in\SSTab(\lambda)}\bsx^T, 
\label{Schur-def}
\eeq
where $\SSTab(\lambda)$ denotes the set of 
all semi-standard tableaux of shape $\lambda$. 
By definition, a semi-standard tableau of shape $\lambda$ 
is an array $T$ of positive integers $T(i,j)$ that are put 
on the cells $(i,j) \in \lambda$ of the Young diagram 
and increasing in the rows and strictly increasing in the columns: 
\beqnn
  T(i+1,j) > T(i,j) \le T(i,j+1). 
\eeqnn
The summand $\bsx^T$ is the monomial 
\beqnn
  \bsx^T = \prod_{(i,j)\in\lambda}x_{T(i,j)} 
\eeqnn
determined by the entries of $T$.  In the same sense, 
the skew Schur function $s_{\lambda/\mu}(\bsx)$ 
is defined by the sum 
\beq
  s_{\lambda/\mu}(\bsx) = \sum_{T\in\SSTab(\lambda/\mu)}\bsx^T 
\label{skew-Schur-def}
\eeq
over the set $\SSTab(\lambda/\mu)$ of all semi-standard tableaux 
of shape $\lambda/\mu$.  The summand is again a monomial 
of the form 
\beqnn
  \bsx^T = \prod_{(i,j)\in\lambda/\mu}x_{T(i,j)}. 
\eeqnn
When $\lambda \not\supset \mu$, we define $s_{\lambda/\mu}(\bsx) = 0$.

In the case of $N$-variables, the entries of tableaux  
are restricted to $[N] = \{1,2,\ldots,N\}$.  
Consequently, the Schur functions for partitions of length 
greater than $N$ vanish, 
\beq
  s_\lambda(x_1,\ldots,x_N) = 0 \quad 
  \text{if $l(\lambda) > N$}, 
\eeq
and the non-vanishing ones are given by the finite sum 
\beq
  s_\lambda(x_1,\ldots,x_N) 
  = s_\lambda(x_1,\ldots,x_N,0,0,\ldots) 
  = \sum_{T\in\SSTab(\lambda,[N])}\bsx^T 
\eeq
over the set $\SSTab(\lambda,[N])$ of semi-standard tableaux 
of shape $\lambda$ with entries in  $[N]$.  

The Schur and skew Schur functions are symmetric functions. 
This fact, far from being obvious in the foregoing definitions 
(\ref{Schur-def}) and (\ref{skew-Schur-def}),  
becomes manifest in the Jacobi-Trudi formulae 
\beq
  s_{\lambda/\mu}(\bsx) 
  = \det\left(h_{\lambda_i-\mu_j-i+j}(\bsx)\right)_{i,j=1}^n 
  = \det\left(e_{\tp{\lambda}_i-\tp{\mu}_j-i+j}(\bsx)\right)_{i,j=1}^m, 
\eeq
where $n$ and $m$ are chosen to be such that 
$n \ge l(\lambda)$ and $m \ge l(\tp{\lambda})$.  
The entries $h_k(\bsx)$ and $e_k(\bsx)$, $k = 0,1,2,\ldots$, 
of the determinants are the complete and elementary symmetric functions 
\beqnn
\begin{gathered}
  h_k(\bsx) = \sum_{i_1\le\cdots\le i_k}x_{i_1}\cdots x_{i_k}, \quad 
  e_k(\bsx) = \sum_{i_1<\cdots<i_k}x_{i_1}\cdots x_{i_k} \quad 
  \text{for $k \ge 1$},\\
  h_0(\bsx) = e_0(\bsx) = 1. 
\end{gathered}
\eeqnn

\subsection{Cauchy identities}

The Schur functions satisfy the Cauchy identities 
\beq
\begin{gathered}
  \sum_{\lambda\in\calP}s_\lambda(\bsx)s_\lambda(\bsy) 
  = \prod_{i,j\ge 1}(1 - x_iy_j)^{-1},\\
  \sum_{\lambda\in\calP}s_\lambda(\bsx)s_{\tp{\lambda}}(\bsy)
  = \prod_{i,j\ge 1}(1 + x_iy_j), 
\end{gathered}
\label{Cauchy-identity}
\eeq
where $\calP$ denotes the set of all partitions.  
These identities are generalized to the skew Schur functions as 
\beq
\begin{gathered}
  \sum_{\lambda\in\calP}s_{\lambda/\mu}(\bsx)s_{\lambda/\nu}(\bsy) 
  = \prod_{i,j\ge 1}(1 - x_iy_j)^{-1}
    \sum_{\lambda\in\calP}s_{\mu/\lambda}(\bsy)s_{\nu/\lambda}(\bsx),\\
  \sum_{\lambda\in\calP}s_{\lambda/\mu}(\bsx)s_{\tp{\lambda}/\tp{\nu}}(\bsy)
  = \prod_{i,j\ge 1}(1 + x_iy_j)
    \sum_{\lambda\in\calP}
    s_{\tp{\mu}/\tp{\lambda}}(\bsy)s_{\nu/\lambda}(\bsx). 
\end{gathered}
\label{skew-Cauchy-identity}
\eeq
When $\mu = \nu = \emptyset$, they reduce to (\ref{Cauchy-identity}). 
Moreover, since the Schur and skew Schur functions have the homogeneity 
\beq
  s_\lambda(Q\bsx) = Q^{|\lambda|}s_\lambda(\bsx),\quad 
  s_{\lambda/\mu}(Q\bsx) = Q^{|\lambda|-|\mu|}s_{\lambda/\mu}(\bsx), 
\eeq
one can slightly generalize (\ref{Cauchy-identity}) 
and (\ref{skew-Cauchy-identity}) as 
\beq
\begin{gathered}
  \sum_{\lambda\in\calP}Q^{|\lambda|}s_\lambda(\bsx)s_\lambda(\bsy) 
  = \prod_{i,j\ge 1}(1 - Qx_iy_j)^{-1},\\
  \sum_{\lambda\in\calP}Q^{|\lambda|}s_\lambda(\bsx)s_{\tp{\lambda}}(\bsy)
  = \prod_{i,j\ge 1}(1 + Qx_iy_j), 
\end{gathered}
\label{Cauchy-identity2}
\eeq
and 
\beq
\begin{gathered}
  \sum_{\lambda\in\calP}Q^{|\lambda|}
  s_{\lambda/\mu}(\bsx)s_{\lambda/\nu}(\bsy) 
  = \prod_{i,j\ge 1}(1 - Qx_iy_j)^{-1}
    \sum_{\lambda\in\calP}Q^{|\lambda|}
    s_{\mu/\lambda}(Q\bsy)s_{\nu/\lambda}(Q\bsx),\\
  \sum_{\lambda\in\calP}Q^{|\lambda|}
  s_{\lambda/\mu}(\bsx)s_{\tp{\lambda}/\tp{\nu}}(\bsy)
  = \prod_{i,j\ge 1}(1 + Qx_iy_j)
    \sum_{\lambda\in\calP}Q^{|\lambda|}
    s_{\tp{\mu}/\tp{\lambda}}(Q\bsy)s_{\nu/\lambda}(Q\bsx). 
\end{gathered}
\label{skew-Cauchy-identity2}
\eeq

\section{Topological vertex and partition functions}

In this section, we review the notion of topological vertex 
and the construction of the amplitudes (or partition functions) 
of topological string theory on toric Calabi-Yau 3-folds.  
We refer details to Mari\~{n}o's book \cite{Marino-book} 
and references cited therein, in particular, the original paper 
\cite{AKMV03} of Aganagic et al.

\subsection{Topological vertex }

The topological vertex depends on a common parameter $q$.  
We consider this parameter to be a complex number with $|q| > 1$.  
Relevant generating functions are expanded in negative powers of $q$.  

\begin{figure}
\begin{center}
\includegraphics[scale=1]{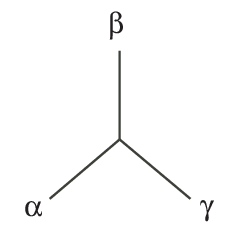}
\end{center}
\caption{Partitions on legs of topological vertex}
\label{fig-vertex}
\end{figure}

The topological vertex has the combinatorial expression 
\beq
  C_{\alpha\beta\gamma}
  = s_{\beta}(q^\rho)q^{\kappa(\gamma)/2}
    \sum_{\nu\in\calP}s_{\alpha/\nu}(q^{\tp{\beta}+\rho})
    s_{\tp{\gamma}/\nu}(q^{\beta+\rho}), 
\label{vertex-def}
\eeq
where $\alpha,\beta,\gamma$ are partitions on the three 
(clockwise ordered) legs emanating from the vertex (Figure \ref{fig-vertex}) 
and $\rho$ is the infinite-dimensional vector 
\beqnn
  \rho = \left(-\frac{1}{2},-\frac{3}{2},\ldots,-i+\frac{1}{2},\ldots\right). 
\eeqnn
The definition of the topological vertex thus contains 
special values of Schur and skew Schur function at 
\beqnn
  q^\rho = \left(q^{-i+1/2}\right)_{i=1}^\infty,\quad 
  q^{\beta+\rho} = \left(q^{\beta_i-i+1/2}\right)_{i=1}^\infty,\quad
  q^{\tp{\beta}+\rho} = \left(q^{\tp{\beta}_i-i+1/2}\right)_{i=1}^\infty. 
\eeqnn
These special values, primarily being power series of $q^{-1}$, 
become rational functions of $q$.  In particular, 
$s_\beta(q^{\rho})$ has the hook formula 
\beq
  s_\beta(q^\rho) 
  = \dfrac{q^{\kappa(\beta)/4}}
      {\prod_{(i,j)\in\beta}(q^{h(i,j)/2}-q^{-h(i,j)/2})}. 
\label{hook-formula}
\eeq
Unfortunately, no hook-like formula seems to be known 
for the special values $s_{\alpha/\nu}(q^{\tp{\beta}+\rho})$ 
and $s_{\tp{\gamma}/\nu}(q^{\beta+\rho})$. 

An extremely nontrivial property of the topological vertex 
is the cyclic symmetry 
\beq
  C_{\alpha\beta\gamma} = C_{\beta\gamma\alpha} = C_{\gamma\alpha\beta}. 
\label{cyclic-symmetry}
\eeq
This property can be derived from a ``crystal model'' \cite{ORV03} 
of the topological vertex.  It seems difficult to prove it directly 
from the conventional knowledge \cite{Macdonald-book} 
on the Schur and skew Schur functions. 
In the special case where $\gamma = \emptyset$, 
the three quantities in (\ref{cyclic-symmetry}) 
can be expressed as 
\beq
\begin{gathered}
  C_{\alpha\beta\emptyset} 
    = s_\beta(q^\rho)s_\alpha(q^{\tp{\beta}+\rho}),\quad 
  C_{\emptyset\alpha\beta}
    = s_\alpha(q^\rho)q^{\kappa(\beta)/2}s_{\tp{\beta}}(q^{\alpha+\rho}),\\
  C_{\beta\emptyset\alpha}
    = q^{\kappa(\alpha)/2}\sum_{\nu\in\calP}
      s_{\beta/\nu}(q^\rho)s_{\tp{\alpha}/\nu}(q^\rho). 
\end{gathered}
\label{2-leg-vertex}
\eeq
Replacing $\beta \to \tp{\beta}$ and using 
the hook formula (\ref{hook-formula}), 
one can reduce the cyclic symmetry in this case 
to the identities 
\beq
    s_\alpha(q^\rho)s_\beta(q^{\alpha+\rho}) 
  = q^{(\kappa(\alpha)+\kappa(\beta))/2}
    \sum_{\nu\in\calP}
      s_{\tp{\alpha}/\nu}(q^\rho)s_{\tp{\beta}/\nu}(q^\rho)
  = s_\beta(q^\rho)s_\alpha(q^{\beta+\rho}). 
\label{2-leg-identity}
\eeq
A direct proof of (\ref{2-leg-identity}) can be found 
in Zhou's paper \cite{Zhou0310a}.

\subsection{Gluing vertices}

The amplitude of topological string theory 
on a toric Calabi-Yau 3-fold $X$ is obtained 
by ``gluing'' the vertex weights 
along the internal edges of the web diagram.    
The vertex weight $C_{\alpha\beta\gamma}$ itself 
is the amplitude of the simplest 3-fold $\CC^3$. 

The internal lines, too, are also weighted.  
The $n$-th internal line of the web diagram 
is given the weight $(-Q_n)^{|\alpha_n|}$, 
where $Q_n$ is the so called ``K\"ahler parameter'', 
and $\alpha_n$ is the partition assigned to 
one of the two vertices connected to this internal line.  
The other vertex have $\tp{\alpha}_n$ on the same line, 
because its leg along this line is given 
an opposite orientation therein.  
The negative sign in the weight is also related 
to this inversion of orientation.  

Thus the $n$-th internal line and the two vertices 
on both ends altogether have the weight 
$C_{\alpha_n\beta_n\gamma_n}(-Q_n)^{|\alpha_n|}
C_{\tp{\alpha}_n\beta'_n\gamma'_n}$. 
This weight is further modified by the ``framing factor'' 
$(-1)^{r_n|\alpha_n|}q^{-r_n\kappa(\alpha_n)/2}$, 
where $r_n$ is an integer determined by directional vectors 
$v_n,v'_n$ of the legs carrying $\beta_n,\beta'_n$: 
\beqnn
  r_n = v'_n \wedge v_n = \det(v'_n,v_n). 
\eeqnn
The total amplitude $Z$, which is referred to 
as a ``partition function'' in the following, 
is obtained by summing the product of these weights 
over all possible configuration of partitions 
on the internal lines:  
\beq
  Z = \sum_{\alpha_1,\cdots,\alpha_N\in\calP} \cdots 
       C_{\alpha_n\beta_n\gamma_n}(-Q_n)^{|\alpha_n|}
       (-1)^{r_n|\alpha_n|}q^{-r_n\kappa(\alpha_n)/2}
       C_{\tp{\alpha_n\beta'_n\gamma'_n}}\cdots 
\eeq

\begin{figure}
\begin{center}
\includegraphics[scale=1]{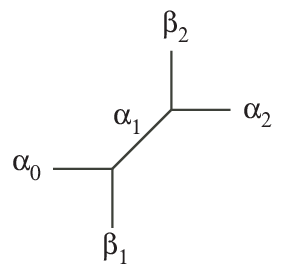}
\end{center}
\caption{Partitions assigned to web diagram of resolved conifold}
\label{fig-conifold}
\end{figure}

Let us illustrate the construction of the partition function 
in the case of the resolved conifold 
$X = \calO(-1)\oplus\calO(-1)\to \CC\PP^1$. 
The web diagram has two vertices as shown in Figure \ref{fig-conifold}.  
We assign the partitions $\alpha_0,\beta_1,\beta_2,\alpha_2$ to 
the external legs and the partition $\alpha_1$ and 
the K\"ahler parameter $Q$ to the internal line.   
The partition function $Z = Z^{\alpha_0\alpha_2}_{\beta_1\beta_2}$ 
is a sum of the form 
\beq
  Z^{\alpha_0\alpha_2}_{\beta_1\beta_2}
  = \sum_{\alpha_1\in\calP}C_{\alpha_1\beta_1\alpha_0}
      (-Q)^{|\alpha_1|}C_{\tp{\alpha_1}\beta_2\alpha_2}. 
\label{Zaa-conifold}
\eeq
(The framing factor vanishes in this case.)  
When $\alpha_0 = \alpha_2 = \emptyset$, the vertex weight 
are simplified to those shown in (\ref{2-leg-vertex}): 
\beqnn
  C_{\alpha_1\beta_1\emptyset} 
    = s_{\beta_1}(q^\rho)s_{\alpha_1}(q^{\tp{\beta_1}+\rho}), \quad 
  C_{\tp{\alpha_1}\beta_2\emptyset}
    = s_{\beta_1}(q^\rho)s_{\tp{\alpha}_1}(q^{\tp{\beta_2}+\rho}). 
\eeqnn
The partition function can be thereby written as 
\beqnn
  Z^{\emptyset\emptyset}_{\beta_1\beta_2}
  = s_{\beta_1}(q^\rho)s_{\beta_2}(q^\rho)
    \sum_{\alpha_1\in\calP}(-Q)^{|\alpha_1|}
    s_{\alpha_1}(q^{\tp{\beta_1}+\rho})
    s_{\tp{\alpha}_1}(q^{\tp{\beta_2}+\rho}), 
\eeqnn
and, by the Cauchy identities (\ref{Cauchy-identity2}), 
boils down to the well known product formula 
\beq
  Z^{\emptyset\emptyset}_{\beta_1\beta_2}
  = s_{\beta_1}(q^\rho)s_{\beta_2}(q^\rho)
    \prod_{i,j=1}^\infty(1 - Qq^{\tp{\beta}_{1,i}+\tp{\beta}_{2,j}-i-j+1}). 
\label{Z00-conifold}
\eeq

\subsection{Implications of cyclic symmetry}

Let us consider implications of the cyclic symmetry 
(\ref{cyclic-symmetry}) of the vertex weights 
in the case of the resolved conifold.  

According to the consequences (\ref{2-leg-vertex}) 
of the cyclic symmetry, the vertex weights 
in the definition (\ref{Zaa-conifold}) 
of $Z^{\emptyset\emptyset}_{\beta_1\beta_2}$ 
have another expression: 
\beqnn
\begin{gathered}
  C_{\alpha_1\beta_1\emptyset}
  = q^{\kappa(\alpha_1)/2}\sum_{\nu_1\in\calP}
      s_{\beta_1/\nu_1}(q^\rho)s_{\tp{\alpha}_1/\nu_1}(q^\rho),\\
  C_{\tp{\alpha}_1\beta_2\emptyset}
  = q^{\kappa(\tp{\alpha}_1)/2}\sum_{\nu_2\in\calP}
      s_{\beta_2/\nu_2}(q^\rho)s_{\alpha_1/\nu_2}(q^\rho). 
\end{gathered}
\eeqnn
Substituting this expression in (\ref{Zaa-conifold}) 
and noting the general property 
\beqnn
  \kappa(\tp{\lambda}) = - \kappa(\lambda)
\eeqnn
of the second Casimir value, one can rewrite 
$Z^{\emptyset\emptyset}_{\beta_1\beta_2}$ 
to the the following triple sum: 
\beqnn
  Z^{\emptyset\emptyset}_{\beta_1\beta_2}
  = \sum_{\alpha_1,\nu_1,\nu_2\in\calP}(-Q)^{|\alpha_1|}
    s_{\beta_1/\nu_1}(q^\rho)s_{\tp{\alpha_1}/\nu_1}(q^\rho)
    s_{\beta_2/\nu_2}(q^\rho)s_{\alpha_1/\nu_2}(q^\rho). 
\eeqnn
By the Cauchy identities (\ref{skew-Cauchy-identity2}) 
for skew Schur functions, the partial sum over $\alpha_1$ 
can be expressed as 
\begin{align*}
 &\sum_{\alpha_1\in\calP}(-Q)^{|\alpha_1|}
  s_{\tp{\alpha_1}/\nu_1}(q^\rho)s_{\alpha_1/\nu_2}(q^\rho) \nonumber\\
 &= \prod_{i,j=1}^\infty(1 - Qq^{-i-j+1}) 
    \sum_{\mu\in\calP}(-Q)^{|\mu|}
    s_{\tp{\nu}_1/\tp{\mu}}(-Qq^\rho)s_{\tp{\nu}_2/\mu}(-Qq^\rho). 
\end{align*}
Thus, up to the pre-factor $\prod_{i,j=1}^\infty(1-Qq^{-i^j+1})$, 
the partition function turns into another triple sum: 
\begin{align*}
  Z^{\emptyset\emptyset}_{\beta_1\beta_2}
  &= \prod_{i,j=1}^\infty(1 - Qq^{-i-j+1})\times \nonumber\\
  &\quad\times 
    \sum_{\mu,\nu_1,\nu_2\in\calP}(-Q)^{|\alpha_1|}
    s_{\beta_1/\nu_1}(q^\rho)s_{\tp{\nu}_1/\tp{\mu}}(-Qq^\rho)
    s_{\beta_2/\nu_2}(q^\rho)s_{\tp{\nu}_2/\mu}(-Qq^\rho).
\end{align*}

Let us note that the pre-factor is essentially the inverse 
of the MacMahon function 
\beq
  M(Q,q) = \prod_{n=1}^\infty(1 - Qq^n)^{-n} 
         = \prod_{i,j=1}^\infty(1 - Qq^{i+j-1})^{-1}
\eeq
with $q$ replaced by $q$.  It is well known 
that the MacMahon function is a generating function 
for weighted enumeration of 3D Young diagrams \cite{ORV03}. 

The last triple sum is also somewhat remarkable, 
because the partial sums over $\nu_1$ and $\nu_2$ 
are special values of the so called ``supersymmetric'' 
skew Schur functions 
\beq
  s_{\lambda/\mu}(\bsx\mid\bsy) 
  = \sum_{\nu\in\calP}s_{\lambda/\nu}(\bsx)s_{\tp{\nu}/\tp{\mu}}(\bsy). 
\eeq
Thus we find another expression of $Z^{\emptyset\emptyset}_{\beta_1\beta_2}$: 
\beq
  Z^{\emptyset\emptyset}_{\beta_1\beta_2}
  = \prod_{i,j=1}^\infty(1 - Qq^{-i-j+1})
    \sum_{\mu\in\calP}(-Q)^{|\mu|}
    s_{\beta_1/\mu}(q^\rho\mid -Qq^\rho)
    s_{\beta_2/\tp{\mu}}(q^\rho\mid -Qq^\rho). 
\label{Z00-conifold-bis}
\eeq

The existence of the two expressions (\ref{Z00-conifold}) 
and (\ref{Z00-conifold-bis}) of the partition function 
$Z^{\emptyset\emptyset}_{\beta_1\beta_2}$ is a special feature 
of the resolved conifold.  In particular, 
this implies the non-trivial identities 
\begin{align}
 &s_{\beta_1}(q^\rho)s_{\beta_2}(q^\rho)
  \prod_{i,j=1}^\infty
  \frac{1-Qq^{\tp{\beta}_{1,i}+\tp{\beta}_{2,j}-i-j+1}}{1-Qq^{-i-j+1}}
 \nonumber \\
 &= \sum_{\mu\in\calP}(-Q)^{|\mu|}
    s_{\beta_1/\mu}(q^\rho\mid -Qq^\rho)
    s_{\beta_2/\tp{\mu}}(q^\rho\mid -Qq^\rho). 
\end{align}

\subsection{``On-strip'' case}

\begin{figure}
\begin{center}
\includegraphics[scale=1]{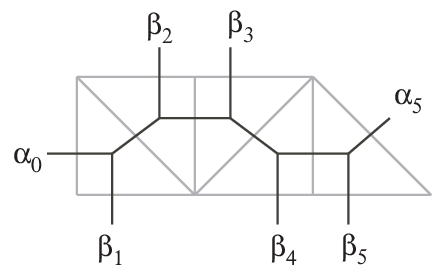}
\end{center}
\caption{Web diagram on triangulated strip}
\label{fig-gen-conifold}
\end{figure}

The product formula (\ref{Z00-conifold}) 
is extended by Iqbal and Kashani-Poor \cite{IKP04} 
to toric Calabi-Yau 3-folds whose toric diagrams 
are triangulations of a strip (Figure \ref{fig-gen-conifold}). 
The associated web diagram is acyclic, 
and each vertex has a vertical external leg.  
We assign the partitions $\beta_1,\ldots,\beta_N$ 
to these vertical legs, the partitions $\alpha_0,\alpha_N$ 
to the non-vertical legs of the leftmost and rightmost vertices, 
and the partitions $\alpha_1,\ldots,\alpha_N$ 
and the K\"ahler parameters $Q_1,\ldots,Q_N$ to the internal lines. 
The partition function $Z^{\alpha_0\alpha_N}_{\beta_1\ldots\beta_N}$ 
is thus given by a sum with respect to $\alpha_1,\ldots,\alpha_N$. 

Following the notations of Nagao \cite{Nagao09} 
and Su{\l}kowski \cite{Sulkowski09}, 
we now introduce the indices $\sigma_n = \pm 1$, $n = 1,\ldots,N$, 
that represent the ``type'' of the vertices: 
\begin{itemize}
\item $\sigma_n = +1$ if the vertical leg of the $n$-th vertex 
is ``down''. 
\item $\sigma_n = -1$ if the vertical leg of the $n$-th vertex 
is ``up''. 
\end{itemize}
For example, for the web diagram of Figure \ref{fig-gen-conifold}, 
\beqnn
  \sigma_1 = +1,\quad \sigma_2 = -1,\quad \sigma_3 = -1,\quad 
  \sigma_4 = +1,\quad \sigma_5 = +1. 
\eeqnn

With these notations, one can summarize the result        
of Iqbal and Kashani-Poor \cite{IKP04} into 
the following beautiful formula: 
\begin{align}
  &Z^{\emptyset\emptyset}_{\beta_1\cdots\beta_N} \nonumber\\
  &= s_{\beta_1}(q^\rho)\cdots s_{\beta_N}(q^\rho) 
   \prod_{1\le m<n\le N}\prod_{i,j=1}^\infty
   \left(1 - Q_{m,n-1}q^{\tp{\beta^{(m)}}_i + \beta^{(n)}_j -i-j+1}
   \right)^{-\sigma_m\sigma_n}. 
\label{Z00-gen-conifold}
\end{align}
Here we have introduced the abbreviation 
\beqnn
  Q_{m,n} = Q_mQ_{m+1}\cdots Q_n
\eeqnn
and define $\beta^{(n)}$ as 
\beqnn
  \beta^{(n)} 
  = \begin{cases}
    \beta_n & \text{if $\sigma_n = +1$},\\
    \tp{\beta_n}& \text{if $\sigma_n = -1$}. 
    \end{cases}
\eeqnn

The method of Iqbal and Kashani-Poor is based 
on a nested use of the Cauchy identities.  
On the other hand, Nagao \cite{Nagao09} and 
Su{\l}kowski \cite{Sulkowski09}, 
generalizing the work of Eguchi and Kanno \cite{EK03}, 
presented a formula of $Z^{\alpha_0\alpha_N}_{\beta_1\cdots\beta_N}$ 
in terms of free fermions and vertex operators \cite{ORV03,BY08}.  
(\ref{Z00-gen-conifold}) can be derived from the fermionic formula 
as well.  This is also a place where integrable hierarchies 
come into the game, because the same fermions and vertex operators 
are also fundamental tools for integrable hierarchies \cite{MJD-book}.

\section{Generating functions of partition functions}

The partition functions $Z^{\alpha_0\alpha_N}_{\beta_1\cdots\beta_N}$ 
are amplitudes of ``open'' topological string theory.  
The partitions $\alpha_0,\alpha_N,\beta_1,\ldots,\beta_N$ 
represent ``boundary conditions'' to string world sheets.  
One can construct generating functions of these open string amplitudes 
by multiplying them with auxiliary Schur functions and summing 
over all possible configurations of partitions on the legs.  
These generating functions are closely related 
to integrable hierarchies \cite{ADKMV03,Zhou0310b,DZ12} 
and quantum mirror curves \cite{GS11,Zhou12}.

\subsection{Various generating functions}

The partition functions $Z^{\emptyset\emptyset}_{\beta_1,\ldots,\beta_N}$ 
can be packed into the the generating function 
\beq
  Z^{\emptyset\emptyset}(\bsx_1,\ldots,\bsx_N) 
  = \sum_{\beta_1,\ldots,\beta_N\in\calP}
    Z^{\emptyset\emptyset}_{\beta_1,\ldots,\beta_N}
    s_{\beta_1}(\bsx^{(1)})\cdots s_{\beta_N}(\bsx^{(N)}) 
\label{Z00(bsx1,...,bsxN)-def}
\eeq
of the $N$-tuple of variables $\bsx^{(n)} 
= (x^{(1)}_1,x^{(2)}_2,\ldots)$, $n = 1,\ldots,N$.  
This function can be specialized to the generating functions 
\beq
  Z_n(\bsx)
   = Z^{\emptyset\emptyset}(\ldots,\bszero,\underset{(n)}{\bsx},\bszero,\ldots) 
  = \sum_{\beta_n\in\calP}
    Z^{\emptyset\emptyset}_{\ldots,\emptyset,\beta_n,\emptyset,\ldots}
    s_{\beta_n}(\bsx) 
\label{Z_n(bsx)-def}
\eeq
of a single set of variables $\bsx = (x_1,x_2,\ldots)$.  
As we shall see below, $Z_n(\bsx)$'s are tau functions 
of the KP hierarchy \cite{MJD-book} 
with respect to the ``time variables'' 
\beq
  t_k = \frac{1}{k}\sum_{i\ge 1}x_i{}^k, \quad 
  k = 1,2,\ldots, 
\label{KP-times}
\eeq
which are nothing but the so called ``power sums'' 
divided by the degree $k$.  
Presumably, $Z^{\emptyset\emptyset}(\bst^{(1)},\ldots,\bsx^{(N)})$ 
will be a tau function of the ``$N$-component'' KP hierarchy 
with respect to the $N$-tuple of time variables 
\beqnn
  t^{(n)}_k = \frac{1}{k}\sum_{i\ge 1}x^{(n)}_i{}^k, 
  \quad k = 1,2,\ldots,
  \quad n = 1,\ldots,N, 
\eeqnn
though we do not have a proof.  

These are generating functions with the leftmost and rightmost 
partitions $\alpha_0,\alpha_N$ being suppressed to $\emptyset$.  
If these partitions are turned on, a new generating function 
can be obtained: 
\beq
  Z_{\beta_1\cdots\beta_N}(\bsy,\bsz) 
  = \sum_{\alpha_0,\alpha_N\in\calP}
    Z^{\alpha_0\alpha_N}_{\beta_1\cdots\beta_N}
    s_{\alpha_0}(\bsy)s_{\alpha_N}(\bsz).  
\label{Zaa(bsy,bsz)-def}
\eeq
From the fermionic representation of 
Nagao \cite{Nagao09} and Su{\l}kowski \cite{Sulkowski09}, 
one can deduce that this is a tau function of 
the 2-component KP hierarchy or, rather, the Toda hierarchy 
(with the lattice coordinate $s$ fixed to $s = 0$).  
Let us mention that this generating function is similar 
to the tau function in the melting crystal model 
of supersymmetric 5D $U(1)$ gauge theory \cite{NT07,NT08}.  

The most general generating function is, of course, 
obtained by turning on all partitions: 
\begin{align}
 &Z(\bsy,\bsx^{(1)},\ldots,\bsx^{(N)},\bsz) \nonumber\\
 &= \sum_{\alpha_0,\beta_1,\ldots,\beta_N,\alpha_N\in\calP}
    Z^{\alpha_0\alpha_N}_{\beta_1,\ldots,\beta_N}
    s_{\alpha_0}(\bsy)s_{\beta_1}(\bsx^{(1)})
      \cdots s_{\beta_N}(\bsx^{(N)})s_{\alpha_N}(\bsz). 
\end{align}
It is natural to expect that this function, too,  
becomes a tau function of the multi-component KP hierarchy.

\subsection{Generating functions as tau functions}

Let us explain why $Z_n(\bsx)$'s may be thought of 
as tau functions of the KP hierarchy.  
This is based on the fact that the coefficients of $Z_n(\bsx)$ 
take the factorized form 
\beq
\begin{aligned}
  Z^{\emptyset\emptyset}_{\ldots,\emptyset,\beta_n,\emptyset,\ldots}
  &= s_{\beta_n}(q^{\rho})
    \prod_{i=1}^\infty f_{\beta_{n,i}-i+1}
    \prod_{i=1}^\infty g_{\tp{\beta}_{n,i}-i+1}\\
  &\quad\mbox{}\times 
    \prod_{l<m,l\not=n,m\not=n}(1 - Q_{l,m-1}q^{-i-j+1})^{-\sigma_l\sigma_m},
\end{aligned}
\label{Z_n(bsx)-coeff}
\eeq
where 
\beqnn
  f_k = \prod_{m=1}^{n-1}\prod_{j=1}^\infty
        (1 - Q_{m,n-1}q^{k-j})^{-\sigma_m},\quad
  g_k = \prod_{m=n+1}^N\prod_{j=1}^\infty
        (1 - Q_{n,m-1}q^{k-j})^{-\sigma_m} 
\eeqnn
if $\sigma_n = +1$ and 
\beqnn
  f_k = \prod_{m=n+1}^N\prod_{j=1}^\infty
        (1 - Q_{m,n-1}q^{k-j})^{\sigma_m},\quad
  g_k = \prod_{m=1}^{n-1}\prod_{j=1}^\infty
        (1 - Q_{n,m-1}q^{k-j})^{\sigma_m} 
\eeqnn
if $\sigma_n = -1$. 

We first consider the simplified generating function 
\beq
  Z(\bsx) = \sum_{\beta\in\calP}s_{\beta}(q^\rho)s_{\beta}(\bsx). 
\label{Z(bsx)-def}
\eeq
This amounts to letting $Q_m = 0$ for $m = 1,\ldots,N$, in $Z_n(\bsx)$. 
Since $s_{\beta_n}(q^\rho) = C_{\emptyset\beta_n\emptyset}$, 
this is nothing but the generating function for $\CC^3$.  
By the simplest Cauchy identities (\ref{Cauchy-identity}), 
one can rewrite $Z(\bsx)$ to an infinite product of the form 
\beq
  Z(\bsx) = \prod_{i,j=1}^\infty (1 - x_iq^{-j+1/2})^{-1} 
    = \prod_{i=1}^\infty \Phi_{q^{-1}}(x_i), 
\eeq
where $\Phi_q(x)$ is the the quantum dilogarithmic function 
\beq
  \Phi_q(x) 
  = \prod_{j=1}^\infty(1 - xq^{j-/2})^{-1} 
  = \exp\left(\sum_{k=1}^\infty\frac{q^{k/2}x^k}{k(1-q^k)}\right). 
\label{quantum-dilog}
\eeq
From the exponential form of the quantum dilogarithm, 
one can see that $Z(\bsx)$ is an exponential function 
of a linear combination of the KP time variables: 
\beq
  Z(\bsx) = \exp\left(\sum_{k=1}^\infty\frac{q^{-k/2}t_k}{k(1-q^{-k})}\right)
  = \exp\left(\sum_{k=1}^\infty\frac{t_k}{k[k]}\right). 
\eeq
Here we have introduced the notation 
\beq
  [k] = q^{k/2} - q^{-k/2} 
\label{[k]-def}
\eeq
that is commonly used in the literature on the topological vertex.  
Since any exponential function of a linear form 
of the time variables is a (trivial) tau function, 
$Z(x)$ is indeed a tau function.  

We now return to $Z_n(\bsx)$.  
The coefficients (\ref{Z_n(bsx)-coeff}) 
of its Schur function expansion are obtained 
from those of $Z(\bsx)$ by multiplying 
$\prod_{i=1}^\infty f_{\lambda_i-i+1}g_{\tp{\lambda}_i-i+1}$. 
It is known in the theory of integrable hierarchies \cite{MJD-book} 
that this is a very special type of transformations 
on the space of tau functions.  The tau functions 
of the KP hierarchy in general have Schur function expansions 
\beqnn
  \tau(\bsx) = \sum_{\lambda\in\calP}a_\lambda s_\lambda(\bsx) 
\eeqnn
in which the coefficients $a_\lambda$ are Pl\"ucker coordinates 
of an infinite dimensional Grassmann manifold 
(the so called ``Sato Grassmannian'').   
The action of $\mathrm{GL}(\infty)$ on this manifold 
induces transformations of tau functions.  
In particular, diagonal transformations 
\beqnn
  a_\lambda \mapsto 
  a_\lambda\prod_{i=1}^\infty f_{\lambda_i-i+1}
  \prod_{i=1}^\infty g_{\tp{\lambda}_i-i+1} 
  \times\text{constant}
\eeqnn
of the Pl\"ucker coordinates are realized by the action 
of diagonal matrices.  Being derived from the (trivial) 
tau function $Z(\bsx)$ by these transformations, 
$Z_n(\bsx)$, too, is a tau function.  

As regards the more universal generating function 
$Z^{\emptyset\emptyset}(\bsx^{(1)},\ldots,\bsx^{(N)})$, 
we are still unble to prove that this is a tau function 
of the $N$-component KP hierarchy.  
In this respect, the case of the resolved conifold 
is rather special and resemble the generating function 
(\ref{Z(bsx)-def}) for $\CC^3$.  Let us specify this case.  

Recall that the partition function of the resolved conifold 
has another expression (\ref{Z00-conifold-bis}).  
By a nested use of the Cauchy identities (\ref{skew-Cauchy-identity2}) 
for skew Schur functions, one can derive from this expression 
the following product formula: 
\begin{align}
 & Z^{\emptyset\emptyset}(\bsx^{(1)},\bsx^{(2)}) \nonumber\\
 & = \prod_{i,j=1}^\infty
    \frac{(1-Qq^{-i-j+1})(1-Qx^{(1)}_iq^{-j+1/2})(1-Qx^{(2)}_jq^{-i+1/2})
          (1-Qx^{(1)}_ix^{(2)}_j)}
         {(1-x^{(1)}_iq^{-j+1/2})(1-x^{(2)}_jq^{-i+1/2})}. 
\label{Z00(bsx1,bsx2)-con1}
\end{align}
This formula can be further converted to the exponential form 
\begin{align}
 & Z^{\emptyset\emptyset}(\bsx^{(1)},\bsx^{(2)}) \nonumber\\
 & = \prod_{i,j=1}^\infty(1 - Qq^{-i-j+1}) 
     \exp\left(
       \sum_{k=1}^\infty\frac{(1-Q^k)(t^{(1)}_k + t^{(2)}_k)}{k[k]}
       - \sum_{k=1}^\infty kQ^kt^{(1)}_kt^{(2)}_k 
     \right). 
\label{Z00(bsx1,bsx2)-con2}
\end{align}
This expression shows that 
$Z^{\emptyset\emptyset}(\bsx^{(1)},\bsx^{(2)})$ is a tau function 
of the Toda hierarchy (hence, of the 2-component KP hierarchy) 
that is independent of the lattice coordinate $s$.  
Actually, this is an ``almost trivial'' tau function 
of the Toda hierarchy,
just as (\ref{Z(bsx)-def}) is a trivial tau function 
of the KP hierarchy.  It is remarkable that such tau functions 
have a non-trivial structure in the $\bsx$ variables.

\subsection{Wave functions as Baker-Akhiezer function}

We now specialize the variables in $Z_n(\bsx)$ 
to $\bsx = (x,0,0,\ldots)$.  Since 
\beq
  s_\lambda(x,0,0,\ldots) 
  = \begin{cases}
    x^k &\text{if $\lambda = (k)$, $k = 0,1,2,\ldots$},\\
    0   &\text{otherwise}, 
    \end{cases}
\eeq
the partition $\beta_n$ in the definition (\ref{Z_n(bsx)-def}) 
of $Z_n(\bsx)$ is restricted to 
\beqnn
  \beta_n = (k),\quad k = 0,1,2,\ldots. 
\eeqnn
Thus $Z_n(\bsx)$ reduces to 
\beqnn
  Z_n(x) = \sum_{k=0}^\infty
    Z^{\emptyset\emptyset}_{\ldots,\emptyset,(k),\emptyset,\ldots}x^k. 
\eeqnn

Let $\Phi_n(x)$ denote the normalized generating function $Z_n(x)/Z_n(0)$, 
namely, 
\beq
  \Phi_n(x) = 1 + \sum_{k=1}^\infty a_kx^k, \quad 
  a_k = \frac{Z^{\emptyset\emptyset}_{\ldots,\emptyset,(k),\emptyset,\ldots}}
    {Z^{\emptyset\emptyset}_{\ldots,\emptyset,\emptyset,\emptyset,\ldots}}. 
\label{Phi_n-def}
\eeq
This function is studied by Kashani-Poor \cite{KP06} 
as a ``wave function'' in topological string theory.  
Actually, this function amounts to the ``dual'' Baker-Akhiezer function 
of the KP hierarchy.  The genuine Baker-Akhiezer function 
corresponds to the generating function 
\beq
  \Psi_n(x) = 1  + \sum_{k=1}^\infty b_k(-x)^k, \quad 
  b_k = \frac{Z^{\emptyset\emptyset}_{\ldots,\emptyset,(1^k),\emptyset,\ldots}}
    {Z^{\emptyset\emptyset}_{\ldots,\emptyset,\emptyset,\emptyset,\ldots}}. 
\label{Psi_n-def}
\eeq
of the partition functions restricted to 
\beqnn
  \beta_n = (1^k) = (\underbrace{1,\ldots,1}_{k}). 
\eeqnn
The sign factor $(-1)^k$ is inserted for matching 
with the usual definition of the Baker-Akhiezer function 
\cite{MJD-book}.  In the language of free fermions, 
$\Phi_n(x)$ and $\Psi_n(x)$  correspond to the fermion fields 
$\psi^*(x)$ and $\psi(x)$.  To be more precise, 
the usual Baker-Akhiezer functions depend 
on the time variables $\bst = (t_1,t_2,\ldots)$ 
of the KP hierarchy; these time variables are now specialized 
to $\bst = \bszero$.

\subsection{$q$-difference equations for wave functions}

Kashani-Poor \cite{KP06} pointed out, in the case of 
the resolved conifold, that these ``wave functions'' 
satisfy linear $q$-difference equations.  
Gukov and Su{\l}kowski \cite{GS11} interpreted 
these equations as a realization of the ``quantum mirror curve'' 
of the resolved conifold.  

One can derive those $q$-difference equations 
for the generalized conifolds as well.  
Since the cases of $\Phi_n(x)$ and $\Psi_n(x)$ are parallel, 
let us explain the derivation for $\Phi_n(x)$ in detail.  

The first thing to do is to express the coefficients 
of this power series in an explicit form.  
To this end, apply the formula (\ref{Z00-gen-conifold}) 
to the case where 
\beqnn
  \beta_n = (k), \quad 
  \beta_m = \emptyset \quad \text{for $m \not= n$}. 
\eeqnn
The hook formula (\ref{hook-formula}) implies 
that $s_{\beta_n}(q^\rho) = s_{(k)}(q^\rho)$ can be expressed as 
\beqnn
  s_{(k)}(q^\rho) = \frac{q^{k(k-1)/4}}{[1][2]\cdots[k]}, 
\eeqnn
recall the definition (\ref{[k]-def}) of $[k]$. 
Thus, after some more algebra, the following expression 
of the coefficients $a_k$ for $k > 0$ can be obtained: 
\begin{align}
   a_k 
   &= \frac{q^{k(k-1)/4}}{[1]\cdots[k]}\prod_{1\le m<n}\prod_{i=1}^k
     (1 - Q_{m,n-1}q^{\sigma_n(k-i)})^{-\sigma_m\sigma_n}\times
   \nonumber\\
   &\quad\times
    \prod_{n<m\le N}\prod_{i=1}^k
    (1 - Q_{n,m-1}q^{-\sigma_n(k-i)})^{-\sigma_n\sigma_m}. 
\end{align}

One can rewrite this somewhat complicated expression 
of $a_k$'s as 
\beq
  a_k = \frac{q^{k(k-1)/4}}{[1]\cdots[k]}
    \frac{C_n(1)C_n(q)\cdots C_n(q^{k-1})}{B_n(1)B_n(q)\cdots B_n(q^{k-1})}, 
\eeq
where $B_n(y)$ and $C_n(y)$ are Laurent polynomials of $y$: 
\beqnn
\begin{gathered}
  B_n(y) 
  = \prod_{1\le m<n,\;\sigma_m\sigma_n>0}(1 - Q_{m,n-1}y^{\sigma_n})
    \prod_{n<m\le N,\;\sigma_m\sigma_n>0}(1 - Q_{n,m-1}y^{-\sigma_n}),\\
  C_n(y) 
  = \prod_{1\le m<n,\;\sigma_m\sigma_n<0}(1 - Q_{m,n-1}y^{\sigma_n})
    \prod_{n<m\le N,\;\sigma_m\sigma_n<0}(1 - Q_{n,m-1}y^{-\sigma_n}). 
\end{gathered}
\eeqnn
Thus $a_k$'s turn out to satisfy the recurrence relations 
\beq
  a_k = a_{k-1}\frac{q^{(k-1)/2}C_n(q^{k-1})}{[k]B_n(q^{k-1})}.  
\eeq
One can thereby derive the $q$-difference equation 
\beq
  [x\rd_x]\Phi_n(x) 
  = x\frac{C_n(q^{x\rd_x})}{B_n(q^{x\rd_x})}q^{x\rd_x/2}\Phi_n(x). 
\label{q-diffeq-Phi}
\eeq
Note that $q^{x\rd_x}$ and $[x\rd_x]$, where $\rd_x = \rd/\rd x$, 
act as $q$-shift and $q$-difference operators: 
\beqnn
  q^{x\rd_x}f(x) = f(qx), \quad
  [x\rd_x]f(x)  = (q^{x\rd_x/2}-q^{-x\rd_x/2})f(x) 
    = f(q^{1/2}x) - f(q^{-1/2}x). 
\eeqnn
(\ref{q-diffeq-Phi}) shows a precise form 
of the $q$-difference equations conjectured 
by Gukov and Su{\l}kowski in a vague form. 

In much the same way, the following $q$-difference equation 
for $\Psi_n(x)$ can be derived: 
\beq
  [-x\rd_x]\Psi_n(x) 
  = x\frac{C_n(q^{-x\rd_x})}{B_n(q^{-x\rd_x})}q^{-x\rd_x/2}\Psi_n(x). 
\label{q-diffeq-Psi}
\eeq
Note that this equation is formally related 
to (\ref{q-diffeq-Phi}) by the inversion $q \to q^{-1}$ 
of the parameter $q$.  

Let us illustrate the $q$-difference equations 
in the case of the resolved conifold.  
$\Phi_1(x)$ and $\Phi_2(x)$ in this case are identical 
and become a power series of the form 
\beq
  \Phi(x) = 1 
  + \sum_{k=1}^\infty
    \frac{q^{k(k-1)/4}(1-Q)(1-Qq^{-1})\cdots(1-Qq^{1-k})}{[1][2]\cdots[k]}x^k. 
\label{Phi-conifold}
\eeq
This power series satisfies the $q$-difference equation 
\beq
  \Phi(q^{1/2}x) - \Phi(q^{-1/2}x)  = x (1 - Qq^{-x\rd_x})\Phi(q^{1/2}x), 
\label{q-diffeq-conifold}
\eeq
which can be rewritten as 
\beqnn
  \Phi(x) = \frac{1 - Qq^{-1/2}x}{1 - q^{-1/2}x}\Phi(q^{-1}x). 
\eeqnn
The last equation implies that that $\Phi(x)$ 
is an infinite product of the form 
\beq
  \Phi(x) = \prod_{n=1}^\infty\frac{1 - Qq^{-n+1/2}x}{1 - q^{-n+1/2}x}, 
\label{Phi-conifold-bis}
\eeq
hence a quotient of two quantum dilogarithmic functions 
defined in (\ref{quantum-dilog}) with $q$ being replaced by $q^{-1}$. 

One can consider the case of $\CC^3$ as the ``decoupling'' limit 
letting $Q \to 0$.  The foregoing results (\ref{Phi-conifold}), 
(\ref{q-diffeq-conifold}) and (\ref{Phi-conifold-bis}) 
thereby reduce to the wave function 
\beq
  \Phi(x) = 1 
  + \sum_{k=1}^\infty\frac{q^{k(k-1)/4}}{[1][2]\cdots[k]}x^k, 
\eeq
the $q$-difference equation 
\beq
  \Phi(q^{1/2}x) - \Phi(q^{-1/2}x)  = x \Phi(q^{1/2}x), 
\label{q-diffeq-C3}
\eeq
and the infinite product formula 
\beq
  \Phi(x) = \prod_{n=1}^\infty (1 - q^{-n+1/2}x)^{-1} 
\label{Phi-C3-bis}
\eeq
for $\CC^3$.  Thus the wave function in this case 
is the quantum dilogarithmic function itself. 

Let us stress that the infinite product formulae 
(\ref{Phi-conifold-bis}) and (\ref{Phi-C3-bis}) 
of the wave functions, which are well known to experts, 
are a special feature of the resolved conifold and $\CC^3$. 
This feature stems from the simple structure 
of the $q$-difference equations (\ref{q-diffeq-conifold}) 
and (\ref{q-diffeq-C3}).  The $q$-difference equation 
(\ref{q-diffeq-Psi}) for other generalized conifolds 
are more complicated, and presumably do not imply 
an infinite product formula of solutions.  

In the classical ($q \to 1$) limit, 
the $q$-difference equation (\ref{q-diffeq-Phi}) 
for $\Phi_n(x)$ turns into the equation 
\beq
  y^{1/2}-y^{-1/2} = x\frac{C_n(y)}{B_n(y)}y^{1/2} 
\label{mirror-curve-eq1}
\eeq
of the ``mirror curve''.   The $q$-difference equation 
(\ref{q-diffeq-Psi}) for $\Psi_n(x)$ yields 
the same equation with $y$ replaced by $y^{-1}$.  
One can rewrite this equation in the form 
\beq
  x = \frac{(1-y^{-1})B_n(y)}{C_n(y)}, 
\label{mirror-curve-eq2}
\eeq
which almost agrees with the one conjectured by Gukov and Su{\l}kowski.  
For example, this equation for the resolved conifold reads 
\beq
  x = \frac{1 - y^{-1}}{1 - Qy^{-1}}. 
\eeq
Although this equations is slightly different from 
the usual mirror curve, the discrepancy can be resolved 
by ``framing transformations'' \cite{GS11,Zhou12}.

\section{Conclusion}

The partition functions $Z^{\alpha_0\alpha_N}_{\beta_1\cdots\beta_N}$ 
of topological string theory on a generalized conifold  
have rich mathematical contents.   In this paper, 
we have mostly considered the case 
where the partitions $\alpha_0,\alpha_N$ 
on the leftmost and rightmost external legs of the web diagram 
are specialized to $\alpha_0 = \alpha_N = \emptyset$.  

Armed with the explicit formula (\ref{Z00-gen-conifold}) 
of these partition functions and the Cauchy identities, 
we have shown the following facts:  
\begin{itemize}
\item The generating functions $Z_n(\bsx)$ 
defined in (\ref{Z_n(bsx)-def}) are tau functions 
of the KP hierarchy.  The $\bsx$ variables are linked 
with the KP time variables $\bst = (t_1,t_2,\ldots)$ 
as shown in (\ref{KP-times}).  
This is a piece of evidence indicating that 
the generating function 
$Z^{\emptyset\emptyset}(\bsx^{(1)},\ldots,\bsx^{(N)})$ 
of $Z^{\emptyset\emptyset}_{\beta_1\cdots\beta_N}$ 
will be a tau function of the $N$-component KP hierarchy. 
\item The wave functions $\Phi_n(x)$ and $\Psi_n(x)$ 
defined in (\ref{Phi_n-def}) and (\ref{Psi_n-def}) 
are the dual pair of Baker-Akhiezer functions 
of the KP hierarchy specialized to $\bst = \bszero$.  
These wave functions satisfy the $q$-difference equations 
(\ref{q-diffeq-Phi}) and (\ref{q-diffeq-Psi}). 
In the classical limit, these $q$-difference equations 
turn into the equation (\ref{mirror-curve-eq1}) 
or, equivalently, (\ref{mirror-curve-eq2}), 
of the mirror curve of the generalized conifold.  
\item The case of the resolved conifold is very special. 
Because of the special form of the $q$-difference equation 
(\ref{q-diffeq-conifold}), the wave function has 
an infinite product form (\ref{Phi-conifold-bis}).  
Moreover, the generating function 
$Z^{\emptyset\emptyset}(\bsx^{(1)},\bsx^{(2)})$, too, 
can be factorized as shown in (\ref{Z00(bsx1,bsx2)-con1}). 
From this factorized form, 
$Z^{\emptyset\emptyset}(\bsx^{(1)},\bsx^{(2)})$ 
turns out to be a very special tau function 
of the 2-component KP hierarchy.  
This reasoning cannot be applied to generalized conifolds.  
\end{itemize}

When $\alpha_0$ and $\alpha_N$ are turned on, 
we can use the fermionic representation of 
$Z^{\alpha_0\alpha_N}_{\beta_1\cdots\beta_N}$ 
\cite{Nagao09,Sulkowski09} to show that 
the generating function $Z_{\beta_1\cdots\beta_N}(\bsy,\bsz)$ 
defined by (\ref{Zaa(bsy,bsz)-def}) is a tau function 
of the Toda hierarchy.  This link with the Toda hierarchy 
will lead to a new perspective of quantum mirror curves.  
This issue will be reported elsewhere.

\subsection*{Acknowledgements}

This work is partly supported by JSPS Grants-in-Aid for 
Scientific Research No. 21540218 and No. 22540186 
from the Japan Society for the Promotion of Science.

\end{document}